\pdfoutput=1
\documentclass[journal]{vgtc}                % final (journal style)
\ifpdf%                                % if we use pdflatex
  \pdfoutput=1\relax                   % create PDFs from pdfLaTeX
  \usepackage{graphicx}                % allow us to embed graphics files
  \DeclareGraphicsExtensions{.pdf,.png,.jpg,.jpeg} % for pdflatex we expect .pdf, .png, or .jpg files
\else%                                 % else we use pure latex
  \usepackage{graphicx}                % allow us to embed graphics files
  \DeclareGraphicsExtensions{.eps}     % for pure latex we expect eps files
\fi%

\graphicspath{{figures/}{pictures/}{images/}{./}} % where to search for the images

\usepackage{microtype}                 % use micro-typography (slightly more compact, better to read)
\PassOptionsToPackage{warn}{textcomp}  % to address font issues with \textrightarrow
\usepackage{textcomp}                  % use better special symbols
\usepackage{mathptmx}                  % use matching math font
\usepackage{times}                     % we use Times as the main font
\usepackage{cite}                      % needed to automatically sort the references
\usepackage{tabu}                      % only used for the table example
\usepackage{booktabs}                  % only used for the table example
\usepackage{todonotes}
\usepackage{balance}
\usepackage{comment}
\onlineid{0}

\vgtccategory{Research}
\vgtcpapertype{please specify}

\newcommand{\ds}[1]{#1}
\newcommand{\cx}[1]{{\textcolor{black}{{#1}}}}
 
\newcommand{\cn}[1]{{\textcolor{black}{{#1}}}}

\title{A Design Space of Vision Science Methods \\ for Visualization Research}

\author{Madison A. Elliott, Christine Nothelfer, Cindy Xiong, and Danielle Albers Szafir}
\authorfooter{
\item
 Madison A. Elliott is with The University of British Columbia. 
 \newline E-mail: mellio10@psych.ubc.ca.
\item
 Christine Nothelfer is with Northwestern University. 
 \newline Email: cnothelfer@gmail.com.
\item
 Cindy Xiong is with the University of Massachusetts Amherst
 \newline E-mail: yaxiong@umass.edu.
\item
 Danielle Albers Szafir is with the University of Colorado Boulder. 
 \newline E-mail: danielle.szafir@colorado.edu.
}

\shortauthortitle{Elliott \MakeLowercase{\textit{et al.}}: Vision Science Methods for Information Visualization}
\abstract{A growing number of efforts aim to understand what people see when using a visualization. These efforts provide scientific grounding to complement design intuitions, leading to more effective visualization practice. However, published visualization research currently reflects a limited set of available methods for understanding how people process visualized data. Alternative methods from vision science offer a rich suite of tools for understanding visualizations, but no curated collection of these methods exists in either perception or visualization research. We introduce a design space of experimental methods for empirically investigating the perceptual processes involved with viewing data visualizations to ultimately inform visualization design guidelines. This paper provides a shared lexicon for facilitating experimental visualization research. We discuss popular experimental paradigms, adjustment types, response types, and dependent measures used in vision science research, rooting each in visualization examples. We then discuss the advantages and limitations of each technique. Researchers can use this design space to create innovative studies and progress scientific understanding of design choices and evaluations in visualization. We highlight a history of collaborative success between visualization and vision science research and advocate for a deeper relationship between the two fields that can elaborate on and extend the methodological design space for understanding visualization and vision. 
} % end of abstract


\teaser{
  \centering
  \includegraphics[width=13cm]{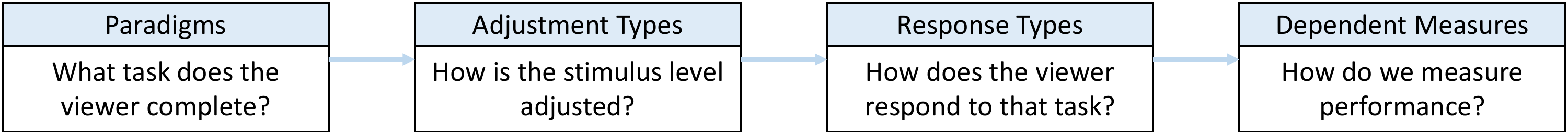}
  \caption{Overview of design space of experimental methods. We present a four component design space to guide researchers in creating visualization studies grounded in vision science research methods. }
  \label{fig:teaser}
}

\vgtcinsertpkg

\begin{document}


\firstsection{Introduction}

\maketitle

%% \section{Introduction} %for journal use above \firstsection{..} instead
%\todo[inline]{A general observation: the first part of the paper focuses on motivating the cross-over between the fields, while the meat of the paper focuses on surveying methods from vision science that might benefit visualization. I might advocate for foregrounding the methods a bit more early in the paper: it's not a hard sell to VIS folks that they need to care (at least a little bit) about vision, but the thing that will get them excited is a catalogue of tools/methods for improving their own research, and, putting my cs hat on, the catalogue here is AWESOME}

Visualization researchers are increasingly interested in running hypothesis-driven empirical studies to investigate human visual perception and intelligence for data displays \cite{rensink_visualization_2014, szafir_four_2016, lam_empirical_2012, kosara_empire_2016}. Efforts such as BeLiV, VISxVISION \cite{nothelfervision}, ETVIS \cite{burch2017eye}, and others continue to promote interdisciplinary science between visualization and experimental psychology. Despite well-established synergy between these fields, there is no shared guide or lexicon for facilitating and designing perceptual visualization experiments. Methodological guides in experimental psychology are either broadly focused \cite{wixted2018stevens} or technique-specific \cite{gescheider2013psychophysics}, 
%and thus, not directly useful for thinking about 
\ds{limiting their utility for} visualization studies. Replete with field-specific jargon, these resources 
%are primarily useful for readers who possess a strong academic background in 
\ds{require extensive knowledge from psychology to interpret effectively. The lack of accessible knowledge about experimental methods for understanding visualizations creates barriers for researchers first seeking to engage in experimental research and limits engagement with a broader suite of tools to understand perception and cognition in visualization}. There is a desire for rigorous experimentation, yet no common ground or established ``basic knowledge'' to facilitate or evaluate it.

Our paper provides a set of tools to afford novel, collaborative research between visualization and vision science researchers \ds{by establishing a preliminary methodological design space for visualization experiments.} \cn{Vision science is the study of how vision works and is used. %While research in this area typically concerns human vision, it can also extend to other animals. 
It is most closely associated with psychology, but draws on a range of disciplines, including cognition and neuroscience.}
%, and computational vision.} %\cn{Our design} space identifies critical components of experimentation, and draws on vision science research to fill the space with popular experimental techniques.
While visualization and vision science research have already seen tangible benefits from collaboration in both design and methodology (e.g., \cite{szafir_modeling_2018, rensink_perception_2010,haroz2012capacity, harrison_ranking_2014, brewer2003colorbrewer, gramazio_colorgorical:_2017, bylinskii_eye_2015}), 
% we see a more general benefit for visualization in 
increasing both the quality and variety of methods used to understand and evaluate visualization design would more generally benefit visualization research. Adapting methods from vision science 
%would be useful for answering questions
\ds{can actionably expand our knowledge} about user attention, memory, visualization design efficacy, and the nature of visual intelligence used to view data. As a first step towards this goal, we catalogue some of the most useful research methods from vision science 
%, with the aim of promoting 
\ds{and discuss their potential for evaluating}
%their use in evaluations of 
user behavior and design for visualizations. 

The two main contributions of this paper are: 1) to provide a \emph{design space} of experimental vision science methods that visualization and perception researchers can use to 
\cn{craft} rigorous experiments and 2) to provide a \cn{shared} \emph{lexicon} \ds{of experimental techniques} to stimulate and engage collaboration between vision and visualization communities. 
%Through these efforts, we 
%\ds{This work} highlights the following opportunities for research in both fields:
%\begin{itemize}
%\item The potential and value of evaluating visualizations with perceptual methods. \vspace{-6pt}
%\item The success of existing interdisciplinary works from vision science and visualization and opportunities for interdisciplinary innovation.\vspace{-6pt}
%\item The most common and useful experimental methods that can be adapted for future scientific collaborations between both fields.
We highlight the potential of using perceptual methods for understanding visualization, discuss prior successes in interdisciplinary research, and identify common methods and opportunities for innovation between the two fields. \ds{Through this effort, w}e aspire to motivate further collaboration and intellectual reciprocity between researchers in both areas of study.

\section{Background}
In this section, we discuss broad advantages and limitations of using vision science research methods for visualization studies. We highlight past interdisciplinary work and several contributions that each field has made to the other. %We also discuss our choice of covering research methods for behavior, as opposed to \ds{other aspects of psychology, such as} neural methods for investigating underlying perceptual mechanisms.

\subsection{Trade-Offs of Vision Science Research Methods}

%The visual system is optimized for transmitting vast amounts of information with minimal effort. For instance, we easily notice {\LARGE large} or \textbf{bold} font in a page of regular text, or a particularly tall bar in a bar chart. This is an example of ``pop out:'' our visual system can process an entire scene (i.e., a page of a paper) in parallel and find key differences immediately \cite{wolfe_guided_1994}. However, readers who have scoured the pages of \textit{Where's Waldo} know that some visual searches are much more difficult and time-consuming than others, requiring planning and a conscious guidance of attention~\cite{palmer2011shapes}. 

Visualizations offload cognitive work to the visual system to help people make sense of data, imparting visual structure to data to, for example, surface key patterns like trends or correlations or identify emergent structures. The amount of empirical research in visualization is increasing \cite{lam_empirical_2012}, and evaluations of user performance are among the top three types of evaluations in published articles. The growing popularity of \cn{quantitative} experiments for visualization evaluation correlates with a movement %within the field
to develop empirical foundations for long-held design intuitions. However, meta-analyses of research methods used in visualization studies have called into question the \ds{reliability of current studies \cite{crisan_how_2018} and even the quality of assumptions from seminal visualization work} \cite{kosara_empire_2016}. Studies have demonstrated that rigorous experimental methods grounded in perceptual theory can expose the limitations of well-accepted conclusions \cite{rensink_perception_2010, szafir_modeling_2018, harrison_ranking_2014}.

The important takeaway from these examples is not that visualization evaluations sometimes get things wrong, but instead that science and our subsequent understanding of the world is constantly evolving. \ds{The methods used to understand these phenomena dictate the efficacy and reliability of that understanding.} Researchers borrowing methods from other disciplines \ds{should do so with an awareness of} the origin and purpose of those methods in the greater context of what they are used to study. Methods and experimental designs evolve with \ds{our knowledge of the world}, which \ds{is in turn} shaped by the progression of theories and evidence. The design of user evaluation and performance studies should not be formulaic and rigid, but rather a creative and thoughtful consideration of the current state of \ds{related} vision science research that guides careful selection of experimental methods \cite{beveridge_art_2004}. \ds{Vision science methods reflect} over a century of work in exploring how people transform real-world scenes and objects into information. This maturity has allowed researchers the time to fail, iterate, improve, converge, and replicate key findings to support well-established theories. 
%For instance, theories of visual attention have evolved considerably even in the past 50 years. \ds{For decades, researchers debated two competing models of visual attention:} Feature Integration Theory \cite{treisman_feature-integration_1980} and Guided Search \cite{wolfe_guided_1994}. Both theories posited some similar ideas (e.g., that salient target objects ``pop out'' and that targets sharing features with distractors is more difficult to find) but deviated on \ds{questions like} what characteristics of a display \ds{influenced people's abilities to find specific information.} Many years and \textit{hundreds} of studies worth of experimental evidence from visual search tasks has led most of the field to advance in embracing Guided Search \cite{wolfe_five_2017}, though \ds{examining these two models using different methods greatly expanded and refined our understanding of visual attention generally}. Studying how visualizations, a more complex visual stimulus than most used in basic vision research, can afford visual search tasks, such as how viewers find the biggest node in a network or how they identify the point of highest sales in a line chart \cite{albers2014task}, enables researchers to validate and refine vision science models, and at the same time, offers important insight and challenges to visualization research and design.

\ds{One concern with adapting vision science to visualization is that vision science focuses on basic research, emphasizing understanding the visual system rather than the functions of different designs. However, design and mechanism are not mutually exclusive: understanding how we see data can drive innovative ideas for how to best communicate different properties of data. For example, understanding the features involved in reading quantities from pie charts can drive novel representations for proportion data \cite{kosara2019circular}. Researchers must carefully consider how the experimental designs can capture different intricacies of visualizations, offering opportunities for methodological innovation balancing control and ecological validity through approaches like applied basic research \cite{shneiderman2016new}.}

\subsection{Past Collaborations and Recent Success}
Visual perception is widely considered a key element of data visualization. 
%Canonical \ds{texts} in the field reflect on classical ideas from vision. For example, Ware \cite{ware_information_2012} \ds{uses ideas from cognition and perception to help readers learn what makes a visualization successful.} Munzner \cite{munzner_visualization_2014} synthesizes classic vision and visualization literature to document an information visualization design space for thinking operationally about how to encode data values. Healey \& Enns \cite{healey2011attention} surveys theories of visual attention in vision science and their application to data visualization.
Two common vision science metrics---\textit{accuracy} and \textit{response time}---have been \ds{widely used} in visualization evaluation studies. While \ds{past methodological adaptation offers insight into effective visualization design,} several studies fall short of their goals due to methodological flaws (see Kosara \cite{kosara_empire_2016} and Crisan \& Elliott \cite{crisan_how_2018} for reviews). For example, early work in graphical perception \cite{cleveland_graphical_1984,lewandowsky_discriminating_1989} suffered from mistakes in precision of design and lack of connection to perceptual mechanisms and often can not be replicated \cite{rensink_perception_2010}, making them less useful for creating design guidelines.

Recent interdisciplinary studies between vision science and visualization \ds{continue to expand our understanding of how and when visualizations work. }\ds{For example, color perception and encoding design is largely informed by experimental methods. These studies offer insights into the most effective choices for encoding design \cite{brewer2003colorbrewer,gramazio_colorgorical:_2017} and application \cite{schloss_mapping_2019}.} \ds{While a full survey of topics in visualization experiments is beyond the scope of this paper, studies have investigated} basic visual features, like orientation \cite{talbot2012empirical}, contrast \cite{mittelstadt2014methods}, grouping \cite{haroz2012capacity}, motion \cite{veras2019saliency}, and redundant use of color and shape \cite{nothelfer2017redundant}. \cx{For example, studying correlation perception in scatterplots using methods from vision science has led to a new understanding of visualization concepts and design \cite{rensink_perception_2010,harrison_ranking_2014,elliott_interference_2016}.} Rensink \& Baldridge \cite{rensink_perception_2010} used psychophysics methods (e.g., see \S\ref{psychophysics}) to derive just noticeable differences (JNDs) \ds{for correlation magnitudes in scatterplots}. These JNDs followed Weber’s law, meaning that \ds{sensitivity to correlation varies predictably and correlation perception is likely a systematic, early (i.e., low-level) visual process and an instance of ensemble coding} \cite{whitney_ensemble_2018, rensink_nature_2017}. This work advances vision science’s understanding of ensemble processing, adding correlation to the types of heuristic information that can be processed rapidly and accurately. Later work replicated and extended these methods to understand how viewers perceive \ds{correlation in} complex displays to inform scatterplot designs \cite{rensink_visualization_2014,harrison_ranking_2014}. % \cx{demonstrating a significant degree of interference for attending to and discriminating} \ds{visually distinct classes in a multi-class scatterplot,} suggesting that complex ensembles are seemingly processed differently than simple objects in attention \cite{elliott_interference_2016}.
\ds{More recent work has leveraged psychophysics models to quantitatively guide visualization design in applications like color encoding design \cite{szafir_modeling_2018} and uncertainty visualization \cite{kale2018hypothetical}.} 

\begin{figure}[t!]
 \centering
 \includegraphics[width=3.5in]{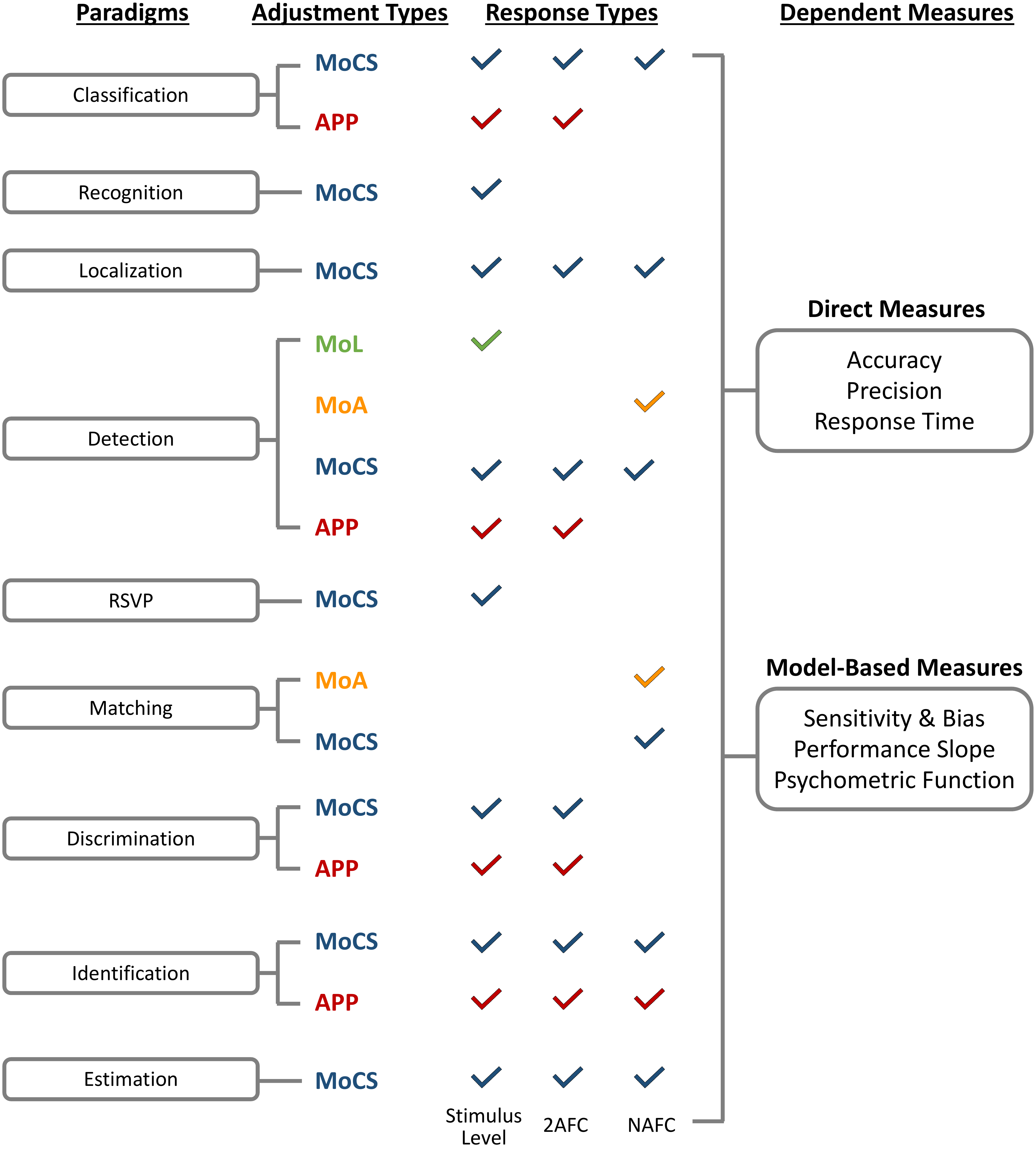}
 \caption{Summary structure of the design space. Starting from left to right, researchers can select a paradigm \cn{and} match it with the appropriate adjustment and response types to obtain the desired dependent measures. \cx{Not all connections between adjustment types and response types are meaningful. Check marks indicate common combinations of adjustment and response types. Adjustments are abbreviated as follows: Method of Adjustment (MoA), Method of Limits (MoL), Method of Constant Stimuli (MoCS), and Adaptive Psychophysical Procedures (APP).}}
 \label{structureOverview}
\end{figure}

These studies show the utility of vision methods for visualization and critically demonstrate that visualizations offer useful opportunities for scientific study. %Simple visualizations are easily controlled stimuli and understanding their structure and function can help vision researchers gain insight into the nature of perceptual mechanisms involved in viewing them. 
The design of visualizations has evolved to work effectively with the human visual system, meaning that they implicitly hold valuable information about how we see and think \cite{rensink_visualization_2014}. Their potential for study is still largely untapped. Excitingly, visualizations will continue to provide novel insight and structural phenomena as \ds{we continue to address the} need to display larger and more complex types of information effectively.

\section{A Design Space of Experimental Methods}
Visualization is an inherently interdisciplinary field. Many of its evaluative practices are derived from those used in human computer interaction, psychology, or sociology \cite{crisan_how_2018} and leverage qualitative, quantitative, or mixed approaches. Because of this diversity, there has been little consensus or standardization of evaluative practices, and no explicit ``handbook'' of visualization evaluation procedures exists. Here, we take an important step by proposing a focused design space of \cx{methods from vision science, a fundamentally empirical field, in hopes of inspiring new perspectives on visualization design and research}.

\subsection{Scope}
The objective of this paper is to explore %one important element of experimental design in visualization--
quantitative behavioral studies of user performance grounded in methodologically-relevant practices from vision science.
%designed to explore how we see the world
We focus on quantitative methods because they support better replicability and generalizability and can connect tasks to designs in well-defined and actionable ways. While neural \cn{(i.e., brain-based)} methods and models are integral for vision science, we do not yet understand how to connect these models to actionable visualization outcomes: visualizations are ultimately created to leverage the visual system and produce optimal viewing \textit{behaviors}\cn{, such as \cx{facilitating data-driven} decisions or gained insights}. For these reasons, neural methods are not included here; however, visualizations may offer interesting scenarios for investigating neural activity. Further, physiological measures such as eye-tracking \cite{blascheck2017visualization} and fNIRs \cite{peck2013using} offer objective biometric insights into visualization use; however, these mechanisms require specialized hardware, experimental design, and nuanced interpretations that are beyond the scope of this work. Finally, existing surveys on related topics such as crowdsourcing \cite{borgo2018information} and statistical analysis \cite{kay2016researcher} provide insight that can augment and influence experimental design, but we focus on broader methodological approaches that can be applied across deployment platforms and \cn{can be} analyzed using a variety of statistical techniques.

\begin{figure}[t!]
 \centering
 \includegraphics[width=3.5in]{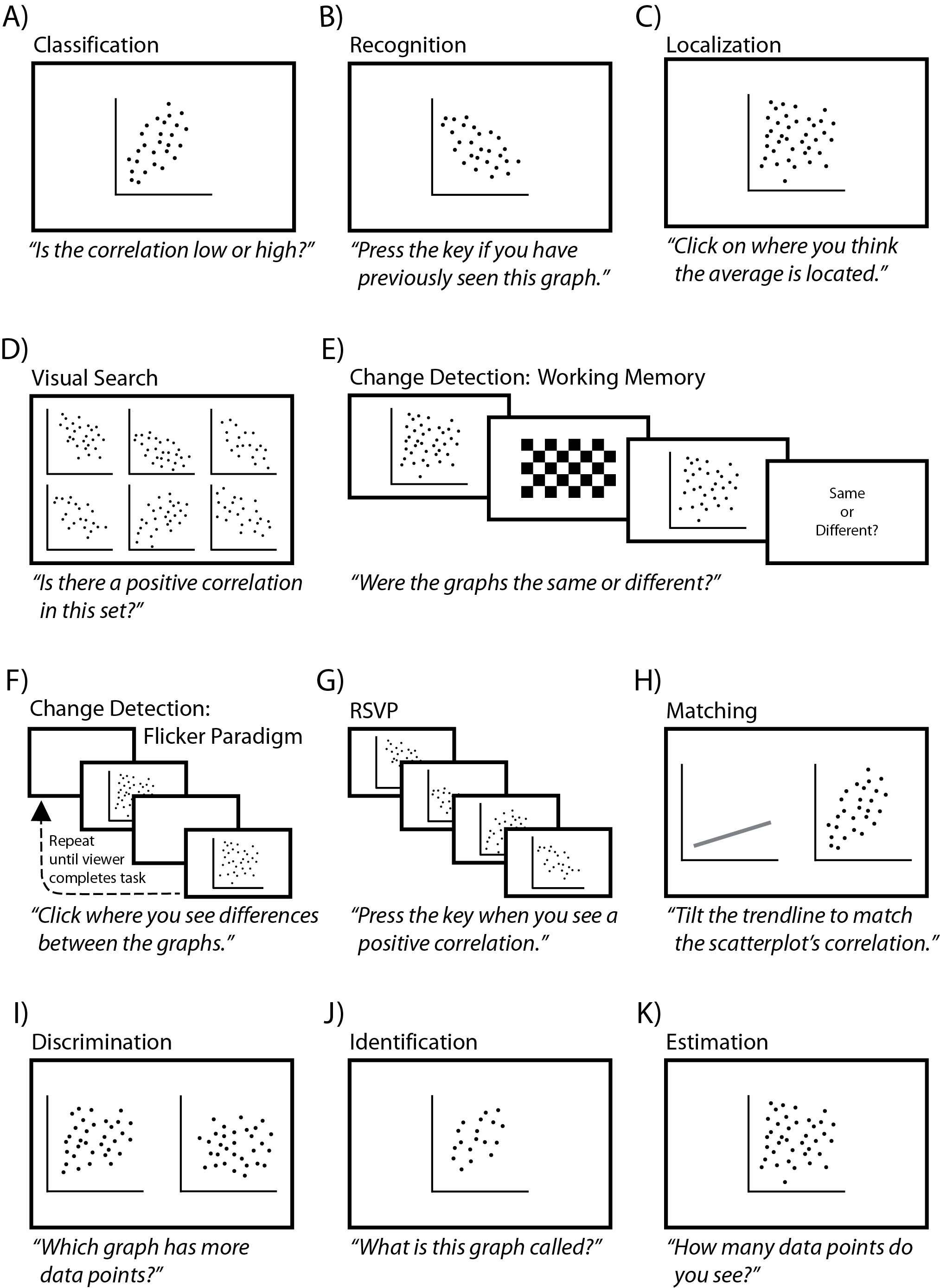}
 \caption{\cx{A running example of scatterplots: eleven ways to study them.} Researchers can ask viewers to categorize what they see (A), whether they recognize a stimulus (B), where a target is located (C), whether a target is present or absent (D), if they remember what changed (E) or if they see the change (F), to indicate when they see a target (G), to match a stimulus to another (H), to compare multiple stimuli (I), to identify what they see (J), or to estimate magnitude (K).}
 \label{paradigms}
\end{figure}

\subsection{Structure of the Design Space}
\ds{Methodological approaches in other fields lie on a spectrum between broad surveys \cite{mcgrath1995methodology,lam_empirical_2012,wixted2018stevens,folstad2012analysis} and specific techniques like research through design \cite{zimmerman2007research} or rapid ethnography \cite{millen2000rapid}. \cx{The design space proposed here balances the complexity of the target problem with the flexibility of a broad survey by organizing relevant experimental techniques \cite{card1997structure,javed2012exploring,diakopoulos2011playable}}. \cn{Design spaces} identify key variables for a design problem, e.g., creating composite visualizations \cite{javed2012exploring}, to provide an actionable structure for systematically reasoning about solutions. While it is tempting to see experimental design as algorithmic and having an ``optimal'' solution, it is a creative process in practice---different experimental approaches to the same research question may yield different results---making it well-suited to design thinking.} %We synthesize key components of the experimental design space and discuss a subset of important instantiations of each element to help structure solutions based on existing approaches and identify opportunities for innovation. Our design space catalogues experimental methods from vision science that can be adapted and used to study how people interpret visualized data. 
\cn{The flexibility of our design space allows researchers to adapt vision science methods for various formats of data collection-- all response types are agnostic of the manner in which viewers provide their response (e.g., speaking, pressing keys, clicking/moving a mouse, writing/typing), and all paradigms could be used as the core task within a neural or physiological study.} Each method is described, connected to the area of vision it was developed to investigate, and then supplemented with relevant concepts in data visualization. To assist in implementing these methods, we also mention relevant analyses and modeling procedures conventionally paired with these methods in order to properly interpret the data and results. % The details of these analysis methods, however, are beyond the scope of this paper.

\label{psychophysics}
We divide \ds{the design space of experimental methods} into four categories (see Figure \ref{fig:teaser} for an overview and Figure \ref{structureOverview} for details) corresponding to key design decisions in crafting a visualization experiment:%\vspace{-6pt}
 \begin{itemize}
     \item \textbf{Paradigm:} What task does the viewer have?\vspace{-8pt}
     \item \textbf{Adjustment Type:} How is the stimulus level adjusted? \vspace{-8pt}
     \item \textbf{Response Type:} How does the viewer respond to that task?\vspace{-8pt}
     \item \textbf{Dependent Measures:} How do we measure performance? 
  
 \end{itemize}
%\noindent By dividing methodological decisions into these phases, researchers can make informed choices based on visualization study trade-offs grounded in existing vision science practices. 

The described methods largely come from \textit{psychophysics}. The basic premise of modern psychophysics is testing a viewer's ability to detect a stimulus, identify a stimulus, and/or differentiate one stimulus from another. \ds{These} methods allow researchers to create descriptive and predictive models of human perception through indirect measurements and probabilistic modeling of responses \cite{farell_psychophysical_1999}.

%%%%%%%%%%%%%%%%%%%%%%%%%%%%%%%%%%%%%%%%%%%%%%%%%%
\section{Paradigms}
Paradigms (Figure \ref{paradigms}) are the specific visual tasks viewers \cn{complete} when using a visualization. Below we detail a set of popular vision science paradigms that may be easily extended to visualization research. Some paradigms are defined by only what the viewer is deciding, while more specialized techniques include additional relevant details such as display specifications, timing, and experimental manipulations.

\subsection{Classification}
\label{classification}
In classification tasks, viewers identify a stimulus by categorizing it---usually according to predetermined choices. Viewers may classify an entire display (e.g., \cn{is the correlation low or high?}) or \cn{specific regions or objects} (e.g., is the target bar in this bar chart larger or smaller than the rest?) \cite{nothelfer2019measures}. %Classification tasks could be a subset of MoCS or Adaptive Psychometric Procedure, depending on how stimuli are selected. Take the above example: %in which a viewer decides whether a scatterplot displays a correlation above or below 0.5 this classification task is considered MoCS if viewers see a pre-set series of scatterplots with correlation values that vary from trial to trial, but would be considered an APP if the scatterplot’s correlation value changed depending on the viewer’s response to the prior trial.\newline

\begin{comment} While these tasks may seem to simply involve the viewer responding with a categorical response, these tasks emphasize categorizing the stimulus based on any given aspect (e.g., Is this graph a scatterplot or a line chart? Is this data group green or blue? Is this correlation positive or negative?). By contrast, tasks where the viewer indicates if/when they have found a certain aspect of a stimulus (as in Visual Search, \S\ref{search}) or whether they do or do not recognize an image (as in Recognition, \S\ref{recog}) would not be considered classification despite viewers providing a categorical response.
\end{comment}

\vspace{4pt}\noindent\textit{\textbf{Advantages:}}
Classification tasks can directly explain how people visually categorize stimuli. Classification is especially useful for both long-term and working memory experiments. A well-designed memory task can be used to disentangle blind guesses, familiar mistakes, and valid/accurate classifications, allowing researchers to quantify what information is retained and confused in memory without having to ask viewers directly (asking directly results in severe bias and noise \cite{rouder2011measure}).

\vspace{4pt}\noindent\textit{\textbf{Limitations:}}
Researchers must effectively determine the ``ground truth" categories of their stimuli. This sometimes requires a pre-test with a training set 
%of stimuli 
or a separate group of viewers rating the test set. Setting categories ahead of time will also bias and constrain viewer responses \cite{livingston1998categorical}, which must be taken into account during data analysis. For this reason, experience with machine learning and/or Signal Detection Theory \cite{emmerich_signal_1967} is useful in designing and analyzing classification tasks.
%both design and analysis of classification tasks.

\subsection{Recognition}
\label{recog}
This paradigm is used to test retention in short and long term memory. Recognition requires the viewer to indicate whether they saw a stimulus previously. A common recognition task involves presenting a set of images, often one at a time, and then later presenting a second set of images composed of ``old'' images (previously seen ones) and ``new'' images. The viewer then indicates whether each image in the second set is old or new. For example, researchers may show a series of scatterplots with varying numbers of data groups and later show a new set of scatterplot groups\cn{, asking} the viewer to indicate which scatterplots appear familiar. Another approach presents the viewer with a constant stream of images and ask them to indicate whenever an image repeats. Recognition tasks have already been used to study memorability in visualizations \cite{borkin_what_2013,borkin_beyond_2016} but are potentially useful for other aspects of visualization perception. %Though recognition is typically a subset of MoCS, it also lends well to non-psychophysical paradigms in which a categorical independent variable is tested. %In these studies, viewers pressed a button when they saw the same visualization repeat in a stream of images, allowing the researchers to uncover what types of visualizations are most memorable.

\vspace{4pt}\noindent\textit{\textbf{Advantages:}}
Recognition tasks are simpler to design and implement (as opposed to classification tasks, for example), and task instructions are easy to explain to viewers. They produce categorical outcomes where the ground truth ``correct" response is known by researchers.

\vspace{4pt}\noindent\textit{\textbf{Limitations:}}
Recognition tasks cannot be used for continuous dependent measures. They typically use 2AFC or NAFC response types, and therefore follow the same limitations (see \S\ref{2AFC}).

\subsection{Localization}
%Many of the paradigms here 
\label{localization}
While classification and recognition are used to understand ‘what’ people see, localization helps us understand ‘where’ people see those items. Localization requires the viewer to indicate the location of a stimulus. It can be used to test different aspects of perception by asking where something is (e.g., attention) or where it previously was located (e.g., memory). Viewers can \textit{click} to indicate where an object is or previously appeared. For example, Smart et al. \cite{smart2019color} asked viewers to click on the mark in a chart (i.e., a data point in a scatterplot, a heatmap square, or a state in a U.S. map) that they thought encoded a particular value. Viewers can also \textit{press a keyboard key} to specify which region of the screen contains the object of interest. For example, Nothelfer et al. \cite{nothelfer2017redundant} briefly showed viewers screens with many shapes and asked viewers to indicate which quadrant of the screen did not contain a particular shape; viewers responded by pressing one of four keys on a number pad, with each key corresponding to a quarter of the screen. Localization studies align well with experiments testing categorical independent variables. 

\vspace{4pt}\noindent\textit{\textbf{Advantages:}}
Localization tasks directly measure spatial attention and also indirectly measure display features and structures that guide or capture attention. Understanding where viewers perceive salient items or structure can inform design by predicting viewer behavior. 

\vspace{4pt}\noindent\textit{\textbf{Limitations:}}
If possible response areas are not explicit (e.g., “click on the box that contained X”), then regions of interest (ROIs) need to be defined to code responses (e.g., any clicks within a 10-pixel radius can be considered a correct localization). Ideally, ROIs are defined \emph{a priori} and spatial overlap is accounted for. Researchers must justify these choices in data cleaning and analysis.

\subsection{Detection}
\label{detection}
Detection requires viewers to indicate whether they perceive
%(d) 
the presence of a particular stimulus. For example, researchers might ask a viewer whether they can \cn{see} %detect grid lines in a chart in order to determine their minimum darkness, or whether they saw 
data points in a scatterplot in order to determine minimum mark size. Detection can be tested when the stimulus is on the screen (i.e., do you detect the target?) or directly after its presentation (i.e., did you detect the target?). Two common types of detection are visual search and change detection.

\subsubsection{Visual Search}
\label{search}
Visual search requires viewers to scan a visual scene to detect a target (object of interest) among distractors (irrelevant objects in the scene). For example, viewers could search for the scatterplot with the highest correlation value in a small multiples display. Search targets can \cn{be} objects that the viewer searches for (e.g., the scatterplot with the highest correlation) or a feature like color (e.g., searching for red dots in a scatterplot with red, blue, and green dots). Visual search has been used to study visual attention for more than 50 years \cite{healey_attention_2012, wolfe_five_2017}.

A basic visual search trial begins with a blank screen showing a small fixation cross. This screen helps control both where the viewer is paying attention (their \emph{spatial attention}) 
and physically looking by restricting the viewer to the same visual starting point prior to each response screen. Next, the stimulus appears on the screen (the \emph{stimulus display onset}). Viewers typically indicate whether a target is “present” or “absent” as quickly as possible, without sacrificing accuracy. Viewers might then also indicate the location of the target if they reported it as present \cite{nothelfer2019measures}.
%Researchers can choose to give accuracy feedback after each trial.

One of the key manipulations in a search task is the number of objects present in the visual search scene (calculated as the number of target(s) + distractor(s)\cn{)}, called \textit{set-size}. Set size is the hallmark of a visual search task and is what distinguishes search from localization. %in which a viewer is simply finding a target (see Localization tasks instead). 
Normally, visual search studies vary the \cn{number} of distractors present over subconditions. For example, if a researcher wanted to know what chart type, connected scatterplot or dual-axis line chart, best facilitates search for positive correlations in a small multiples display, viewers could view small multiples of 5, 10, 15, or 20 connected scatterplots as well as small multiples of 5, 10, 15, or 20 dual-axis line charts. In this case, the study has \cn{four} set-sizes. 
%, and a subcondition of chart type.

Search tasks measure response time (RT) and accuracy. Performance is understood with search slopes (see ``Performance Slopes'' in \S\ref{dep_model}). The slope of the RT $\times$ set size function describes search efficiency. This \cn{function} represents the \textit{search rate}---how much \textit{more} time is required with the addition of each distractor to a visual scene. The steeper the slope, the more time is required to search at larger set sizes, which is called \emph{serial} or \emph{inefficient} search. A flat slope close to a value of 0 indicates “pop out,” meaning that increasing the set size with distracting information does not affect 
%the viewer’s speed of finding 
how quickly people find the target~\cite{wolfe_guided_1994}. %For example, if searching for the dual-axis line chart with the positive correlation yielded a search rate of 20ms, that indicates that the total time it takes to find the positively-correlated chart increases by 20ms every time another chart is added to the small multiples set. 
%The steeper the slope, the more time is required to search at larger set sizes – a behavior described as serial or inefficient search. A flat slope close to a value of 0 indicates “pop out”, meaning that increasing the set size with distracting information does not affect the viewer’s speed of finding the target (Wolfe, 1994).

Visual search studies systematically manipulate set size, and there is normally low error rate across all subconditions \cite{wolfe2008visual}. However, some experimental designs induce higher error rates through brief display times or limiting target rates \cite{wolfe2007low}. Logan \cite{logan_cumulative_2004} provides a 
%good 
review of data modeling techniques for visual search data and their implications for models of attention. Visualization researchers could use search tasks to explore many questions, \cn{such as} which regions of visualizations are salient or important, or how the complexity of a visualization affects what a viewer might attend to.
 %As mentioned above, some visual search tasks may ask the viewer to locate the target (e.g., Which rectangular frame -- left or right -- contains the target? [Nothelfer & Franconeri, VIS 2019])-- these tasks are considered a subset of Localization tasks.

\vspace{4pt}\noindent\textit{\textbf{Advantages:}}
Visual search provides a way to indirectly measure the efficiency of attention. Examining search efficiency could directly inform design guidelines that can help researchers understand how to design more complex displays. Set size manipulations are simple, and the task itself is easy to explain to viewers. % Imagine a visual search task without a set-size manipulation: searching for a positive correlation in 4 dual-axis line charts versus 4 connected scatterplots. Consider that the median response times are 0.7 seconds and 0.8 seconds, respectively, and a statistical test reveals no significant difference between the two chart types. In the absence of manipulating set-size, the researcher cannot examine \textit{efficiency} of visual search via the search slope, which may show that viewers’ search time dramatically increases with set-size for the dual-axis line charts compared to connected scatterplots. Such a finding would yield a wildly different conclusion. 

\vspace{4pt}\noindent\textit{\textbf{Limitations:}}
%Modeling precise accuracy rates is difficult for search tasks that do not also include target localization. 
Search tasks must be designed carefully, and almost always ask participants to localize a detected target (\S\ref{localization}) to rule out random guessing.
%Without location verification, random guessing cannot be ruled out on any given trial. 
Additionally, while efficient search shows that a target captures attention, to generalize the results to design, researchers must determine what is driving this effect, for example, whether it is the mark's physical properties or its contrast with the background and other data.  

\subsubsection{Change Detection}
\label{cd}
Change detection (CD) is used to measure limitations in attention and working memory \textit{capacity}---how much information can be held in mind over a brief interval. CD studies probe whether the viewer was able to notice and remember certain aspects of a stimulus. For example, change detection could help understand what items are attended to in a display and then held in working memory as viewers explore data or view animated transitions.
%UI navigation or large shifts of attention (e.g., scanning multiple charts in a dashboard) is a useful goal for visualization research. 
Two common variations are working memory change detection and the flicker paradigm.

%Much like RSVP (\S \ref{sec:Rapid Serial Visual Presentation (RSVP)}), change detection tasks can be used to study the limits of memory and attention in visualization. Change detection tasks are designed to occur over a much slower time course than RSVP, so they are more useful in the investigation of working memory representation and capacity.

\textit{Working Memory (WM):}
A typical working memory change detection study starts with a blank screen showing a fixation cross, followed by a preliminary display screen showing the “before” stimuli. Often, the “before” display is followed by a \emph{mask} (an unrelated image, like a checkerboard) to disrupt iconic memory 
%(sub-second memory) 
\cite{coltheart1980iconic}, afterimages (such as the bright spot a viewer sees after appearing in a flash photograph) \cite{bradley2012sensory}, and rehearsal strategies \cite{reitman1974without}. After a short delay, an “after” display appears. The “after” display may or may not contain a noticeable change compared to the “before” display. For example, a researcher could show a ``before" display with 5 color-coded clusters of data in a scatterplot. The ``after" display could either show the same 5 colors or change one of the cluster colors, and ask whether viewers detect a difference. Depending on what the 
%researcher is concerned with measuring, 
experiment measures,
the nature of the display and its changes will vary. For example, experiments can test
%may wish to test 
item location, feature information (e.g., color or shape), or direction of motion. In some working memory change detection tasks, researchers record response error distance in physical or feature space (see van den Berg et al. \cite{van_den_berg_variability_2012} for an example). %of an item could change and viewers will be asked to locate where the item came from. In vision science, it is common for the “before” display to several colored squares, and the “after” display to show squares in the same location with either new or old colors. Viewers might be asked to indicate what color a specific square was in the “before” display. 
The design possibilities for WM tasks are broad and largely untapped in visualization. Ma et al. \cite{ma_changing_2014} surveys what WM can tell us about what people see. %In the above example, in which the viewer indicates whether two scatterplots are the same or different, the number of data points may increase/decrease systematically and randoMoLy across trials (Method of Constant Stimuli), or increase/decrease depending on whether the viewer is correct.

\textit{Flicker Paradigm:}
%Another change detection paradigm is the Flicker Paradigm. These 
Tasks in a flicker paradigm continually alternate (``flicker") a display of an original image and a modified image, with a brief blank display in between. The blank display prevents local motion signals from interfering with high-level attentional control \cite{rensink2001change}. Flicker is used to study both working memory as well as change blindness \cite{oregan_change-blindness_1999} (see Rensink \cite{rensink_change_2002} for a survey). Flicker tasks measure the time it takes a viewer to notice the change and their accuracy in identifying the region of the image that has changed.

 %Testing the assumptions and predictions of current models of working memory on simple cases of visualizations in a controlled change detection task could be a promising step towards studying working memory in more complex scenes and displays.

\vspace{4pt}\noindent\textit{\textbf{Advantages:}}
CD tasks are often easy to design and implement, and viewer task instructions are simple to explain. Some tasks, like flicker paradigms, can be as short as a single trial (\emph{one-shot}). One-shot CD tasks can show inattentional blindness and other illusions or robust failures in perception. WM tasks are critical for quantifying memory capacity, and these studies reduce noise and bias due to viewer habituation by making it hard to anticipate when a change occurs in the display.

\vspace{4pt}\noindent\textit{\textbf{Limitations:}}
The biggest limitation in CD tasks lies in modeling the results. There is considerable debate in the working memory research about how to compute \cn{WM} capacity (known as $k$) and how to handle response errors in viewer data~\cite{ma_changing_2014}. This is an active debate among perception researchers and a yet unresolved problem in the field. Analysis and interpretation must be carefully justified by researchers.

\subsection{Rapid Serial Visual Presentation (RSVP)}
RSVP was 
%originally 
designed to explore how viewers comprehend information from a fast series of stimuli \cite{broadbent1987detection}. 
%In this paradigm, 
\cn{This paradigm presents} a set of images, including irrelevant images and at least one target image, in a rapid sequence (see Borkin et al.~\cite{borkin_what_2013,borkin_beyond_2016} for an example). The images are shown one at a time at the same screen location. Viewers identify a target or target category (e.g., ``name the chart type when you see a positive correlation" in a series of different visualizations or ``press the button when you see a positive correlation" in a set of scatterplots).

RSVP experiments manipulate a number of factors, such as timing between stimuli, number of targets present, what type of image precedes a target image, timing between particular irrelevant images and the target image, or timing between target images, known as \emph{lag manipulation}. Lag manipulation is common in RSVP, quantified as the number of irrelevant stimuli which appeared between two target images. For example, if a viewer saw a scatterplot with a positive correlation (the target), then two scatterplots with negative correlations, followed by a positive correlation (the second target), \cn{they will have completed
%this type of trial is called 
a ‘Lag 3’ trial} because the second target appeared 3 images later. 
%RSVP tasks are typically a subset of MoCS, given that various factors may be systematically manipulated.

%RSVP is commonly used to study a phenomenon called attentional blink. Attentional blink describes an occurrence where a viewer fails to notice the second of two targets presented close together in time. It has been studied as an example of attentional limitations, or a suppression of visual processing %Various factors may influence attentional blink such as \cn{x, x, x, and x (need to find citations).}
%This paradigm would be useful for research investigating animation, or displaying changes over time in data, including live, dynamic displays. There is not much visualization research in this area, but past perception research suggests it could be a promising area for future collaborative work \cite{tversky2002animation}.

\vspace{4pt}\noindent\textit{\textbf{Advantages:}}
RSVP tasks are especially useful for examining the time course of attention as well as modeling temporal shifts in attention and their impact on working memory. Well-designed RSVP tasks control eye movements and spatial attention, and would therefore be especially useful for investigating animation or displaying changes over time in data, including live, dynamic displays \cite{tversky2002animation}.

\vspace{4pt}\noindent\textit{\textbf{Limitations:}}
%-Attentional blink, and the accompanying factors that should be considered (though obviously not a limitation if the intention is to learn what visualization-related factors influence the attentional blink!)\newline
RSVP tasks are limited by \cn{an} inability to control gaze fixations due to the nature of the display. As we move our eyes, we miss intermediate visualizations (\emph{saccadic blindness} \cite{raymond_temporary_1992}). One way to mitigate this is by using masking in the stream of images (\S\ref{cd}).

\subsection{Matching}
In matching paradigms, the viewer adjusts one stimulus until it matches another. Viewers can match sub-features of a stimulus (e.g., adjust the luminance of one population in a two-class scatterplot until it matches the other) or match the entire stimuli (e.g., adjust the height of the leftmost bar until it matches the value of the middle bar in a bar chart). Given its versatility, the matching paradigm can be used to study a wide variety of perception and attention topics.

%One interesting application of t
This paradigm is well-suited to \cn{understanding} how well viewers aggregate data across visualization types. For example, experiments might ask viewers to adjust the angle of a trend line until it matches the correlation of the scatterplot \cite{correll2017regression} or to adjust the bar in a bar chart on the left until it matches the mean value of \cn{a} swarm plot on the right. In Nothelfer \& Franconeri \cite{nothelfer2019measures}, viewers adjusted the height of a bar to indicate the average delta in a dual bar chart.

%The matching paradigm could be considered a subset of Method Constant Stimuli or Method of Adjustment. For example, a viewer may adjust a trend line to match the correlation perceived in a scatterplot, and do this for a wide sample of correlation values over the course of the experiment (Method of Constant Stimuli). As an example of Method of Adjustment, a viewer may adjust the contrast of a subset of data points in a scatterplot: some trials may begin with two data groups and ask the viewer to lower the contrast until they see only one data group, while other trials may begin with one data group and ask the viewer to increase the contrast until they see two data groups.

\vspace{4pt}\noindent\textit{\textbf{Advantages:}}
Matching paradigms are optimal for comparing data across visualization types and could be used to evaluate the utility of different design idioms. Matching also indirectly probes whether or not a viewer's mental representation is consistent across designs. 

\vspace{4pt}\noindent\textit{\textbf{Limitations:}}
Matching trials often require unlimited viewing time, so total experiment time could be long. Adjustment methods must be considered carefully (see \S\ref{moa} for details).

\subsection{Discrimination}
\label{discrimination}
In the discrimination paradigm, viewers make comparative judgements about the magnitude of (typically) side-by-side stimuli, such as asking viewers to indicate which of two scatterplots contains more data points. This can be measured at multiple levels of data point numerosity.

Discriminations can be performed across separate stimuli (e.g., two scatterplots on a screen) or within the same stimulus (e.g., two data groups in the same scatterplot). Nothelfer \& Franconeri \cite{nothelfer2019measures} showed viewers dual bar charts, and viewers judged whether there were more increasing or decreasing bar pairs in each display. Rensink \& Baldridge \cite{rensink_perception_2010} asked viewers which of two scatterplots contained a higher correlation---a method extended by Harrison et al. \cite{harrison_ranking_2014} to rank other visualizations of correlation, including parallel coordinates, donut charts, and stacked area charts. Gleicher et al. \cite{gleicher2013perception} asked viewers to indicate which of two data groups in a scatterplot had the higher mean.

%Discrimination studies are typically a subset of Method of Constant Stimuli or Adaptive Psychometric Procedure. In the example above, in which the viewer indicate which scatterplot has more data points, one scatterplot may increase/decrease data points systematically and randoMoLy across trials (Method of Constant Stimuli) or depending on whether the viewer’s prior response was correct (Adaptive Psychometric Procedure).

\vspace{4pt}\noindent\textit{\textbf{Advantages:}}
Discrimination tasks are highly flexible and lend well to adaptive psychometric procedures (see \S\ref{sec:app}). They are the preferred paradigm for evaluating perceptual precision (see \S\ref{dep_direct}) and can be used with complex stimuli such as dashboards.

\vspace{4pt}\noindent\textit{\textbf{Limitations:}}
Potential limitations of discrimination tasks are largely contingent on the Adjustment Type (\S\ref{sec:adj}) used in their implementation. Researchers should be aware that using discrimination to measure accuracy is unnecessarily time-consuming and may be inefficient for subjective measures like preference.
%It should not be the first choice task for studying subjective estimates.

\subsection{Identification}
Identification paradigms require viewers to respond with the identity of the stimulus using open-ended responses. In identification paradigms, experiments typically do not provide a recognition ``template" or training set. Instead, identification tasks are used to study how viewers name stimuli. Viewers can be asked to identify an entire stimulus (e.g., what would you call this chart type?) or to identify a specific feature (e.g., name the color of the less correlated marks in this display).

\vspace{4pt}\noindent\textit{\textbf{Advantages:}}
%As mentioned above, i
Identification paradigms 
%help us understand viewer name stimuli in a less biased manner than classification. T
\cn{offer less biased insight into perceived categories than classification tasks.}
They are also useful when predetermined categories \cn{are unavailable}. %categories are unknown in advance. 
For example, understanding how viewers segment color bins may provide insight on how to design better multihue palettes~\cite{quinan2019examining}. Identification tasks can work as a \emph{pre-task} for classification tasks to generate the categories that are later fed into a classification study. Whereas classification task provide a mental template (categories), identification tasks require viewers to access their individual long term memory store to identify stimuli.

\vspace{4pt}\noindent\textit{\textbf{Limitations:}}
Because viewers are not provided with a mental template or a set of categories, identification trials may take longer than classification trials. Additionally, in some cases researchers should be prepared to account for a wider variety of response categories since they are not constrained ahead of time. This has implications for coding 
%the 
\cn{data} before analysis that must be carefully considered \cite{emmerich_signal_1967}. 

\subsection{Estimation} 
Estimation paradigms require viewers to directly estimate some value of a continuous feature in a display. Magnitude production is the most common type of estimation task, where viewers are required to estimate the magnitude of a stimulus with a numeric response. Estimation tasks are different from classification tasks, which ask viewers to categorize stimuli. For example, if a researcher wished to know how viewers perceive correlation strength in a scatterplot, viewers could \textit{classify} the correlation as "low" or "steep" or \textit{estimate} it at "0.2" or "0.8".
%Experiments could ask a viewer to estimate the number of data points in a scatterplot or to indicate what percentage of counties in a state map voted in favor of a particular proposition. Estimation 

%\cn{While this paradigm may appear similar to the classification paradigm, the critical difference is that viewers \textit{categorize} the stimulus in a classification paradigm, while they \textit{provide a value} on a continuous scale to describe the stimulus in an estimation paradigm. For example, if a researcher wished to know how viewers perceive correlation strength in a scatterplot, viewers could classify the correlation as "low" or "steep", or estimate it at "0.2" or "0.8".}

\vspace{4pt}\noindent\textit{\textbf{Advantages:}}
Estimation tasks measure accuracy, have intuitive instructions, and \cn{are} amenable to various Adjustment Types. To obtain a full psychometric function, the level of the stimulus can be systematically manipulated to understand how close the viewer's response is to its true value at different magnitudes. This function can be used to evaluate, generalize, and predict future estimation performance.

\vspace{4pt}\noindent\textit{\textbf{Limitations:}}
Estimation paradigms do not capture precision (see \S\ref{discrimination}) and should not be used to obtain objective magnitudes from viewers. This is because perceptual estimates of most feature properties are 
%shown to be 
\cn{systematically} biased (e.g., we underestimate mid levels of correlation magnitude \cite{rensink_nature_2017}). \cn{This bias is} often modeled as an instance of Steven's Law, Ekman's Law, or Fechner's Law \cite{teghtsoonian1971exponents}.

%Below are some of the measurement strategies that help researchers do this. 
%%%%%%%%%%%%%%%%%%%%%%%%%%%%%%%%%%%%%%%%%%%%%%%%%%
\section{Adjustment Types}
\label{sec:adj}
\begin{figure}[t]
 \centering
 \includegraphics[width=3.5in]{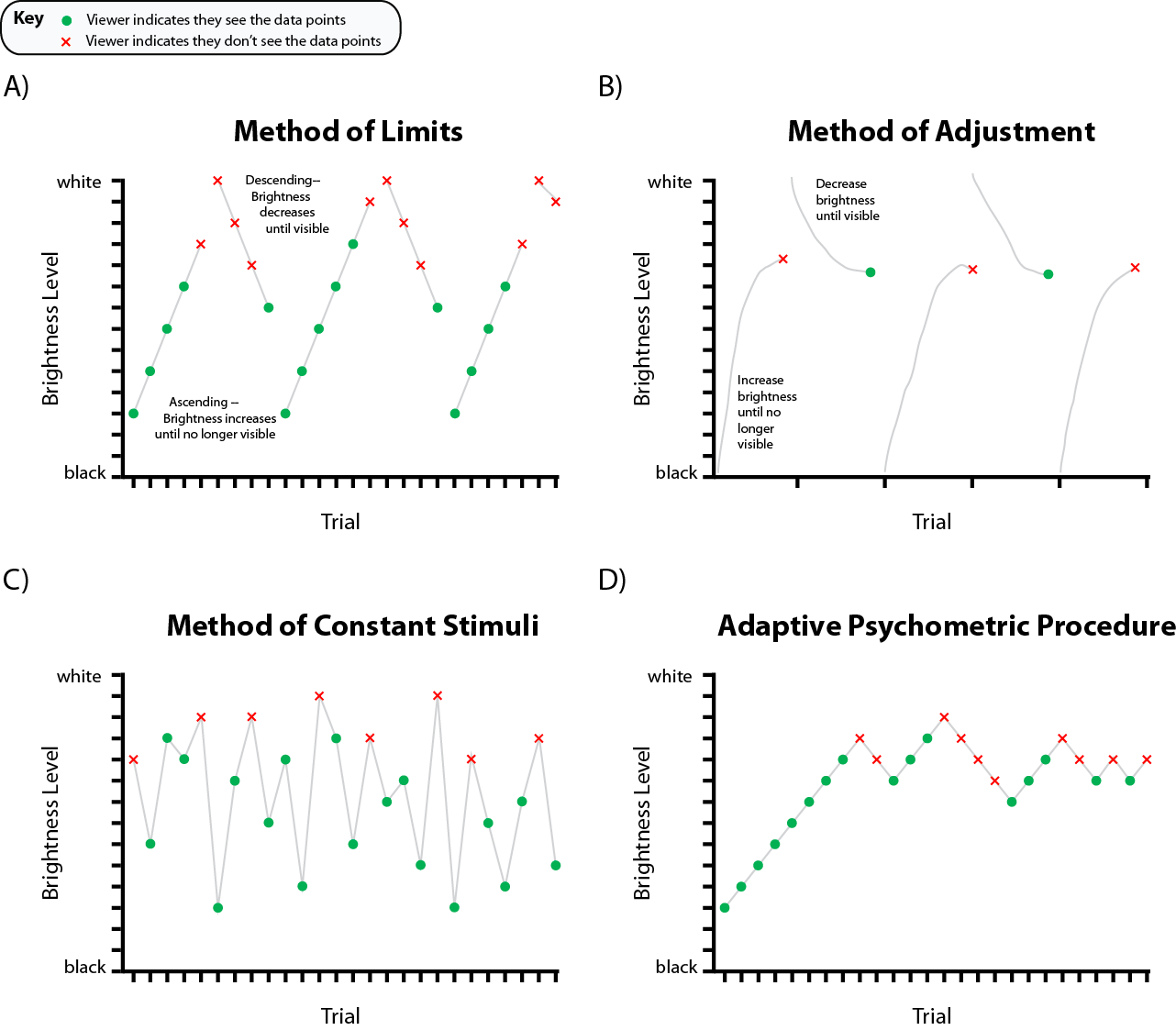}
 \caption{Four adjustment types. Researchers can adjust the brightness of scattered dots on a white background until participants report that they are no longer visible (A), have participants adjust the brightness level of the scattered dots until they are \cn{just} visible (B), present \cn{participants} with random brightness levels and ask them to report whether the dots are visible or not (C), or find \cn{a visibility} threshold by adjusting brightness until the viewer can reliably detect it 75\% of the time (D).}
 \label{methods}
\end{figure}

Psychophysical adjustment types (Figure \ref{methods}) define the overall structure of perceptual experiments by determining the manner in which the stimulus level will be adjusted and responded to. Here, we discuss the three main types---Method of Limits, Method of Adjustment, and Method of Constant Stimuli---and adaptive psychophysical procedures which aid their use.

\subsection{Method of Limits [MoL]}
The goal of the Method of Limits is detection: the researcher wishes to identify the level at which \ds{people see a target property in an image by steadily changing that property until the viewer sees (or no longer sees) the target property.} For example, to detect the upper bound for colors of scatterplot points, an experiment may start with a scatterplot of white dots on a white background and slowly decrease the lightness of the marks until viewers can perceive them. The result gives researchers the highest detectable lightness level that can be used to draw scatterplots. This can be done \ds{using either \emph{ascending} or \emph{descending} methods}, and it is common to use both in a single experiment. Ascending MoL tasks start at a low level of magnitude (often zero) and increase the level of the stimulus over time, requiring viewers to indicate when they can perceive it. Descending MoL tasks start at a high level of the stimulus and decrease its level, requiring viewers to indicate when they can no longer perceive it. \ds{In the scatterplot example above, an ascending MoL design would start with black dots and show viewers increasingly lighter dots until the marks were no longer visible on a white background. A descending design would start with white marks and decrease lightness until viewers report that the marks become visible}. 

\vspace{4pt}\noindent\textit{\textbf{Advantages: }}
%\newline 
	MoL tends to be easy to implement and easy for viewers to understand: studies need to provide viewers with a single value to observe, such as the visibility of marks. These features collectively mean that MoL studies are often fast, affording more trials in a short time. Precisely manipulating a single feature also allows experiments to collect precise measurements about specific phenomena (e.g., color perception) in context using a single stimuli.

\vspace{4pt}\noindent\textit{\textbf{Limitations:}}
	While MoL provides precise per-trial measures, viewers can quickly habituate to trials and often begin predicting when the stimulus will become perceivable or imperceivable. These predictions lead to premature responses, called anticipation errors, that are not precise or accurate representations of perception. Habituated viewers may also become less sensitive to the stimulus overall. Techniques like staircase procedures (\S \ref{sec:app}) help address these limitations in practice.
	
\subsection{Method of Adjustment [MoA]}
\label{moa}
	The Method of Adjustment \ds{operates on the same principles as MoL, but instead of the researcher manipulating a target feature,} 
	%requires viewers to 
	viewers \cn{directly} adjust \ds{properties of a visualization} until they reach a perceptual criteria such as ``present/detectable,'' ``absent/indetectable,'' or ``equal to X.'' This task type is repeated over many trials, and the difference between the correct stimulus level and the viewer response is \ds{typically} recorded and averaged over all trials as a measure of perceptual sensitivity. \ds{In the \cn{MoL} example, viewers would adjust the lightness of scatterplot points until they are just visible.} This trial type could be repeated \ds{for marks with different starting colors to determine an absolute threshold.} \cn{Averaging errors at each level of lightness tested would indicate}
	%could then be averaged to indicate 
 perceptual sensitivity at each level of lightness across different color categories. 
	
\vspace{4pt}\noindent\textit{\textbf{Advantages:}}
	The method of adjustment allows for a broader sampling of \ds{space of possible responses, datasets, and designs} since researchers can vary the distance between the adjustable stimuli \ds{(e.g., the color of a mark)} and \ds{the defined objective (e.g., matching to different colors or backgrounds)} over many trials. The potential for many interleaved conditions and trial types means less risk of habituation and possible increases in statistical power. Viewers can also perform either absolute threshold detection \ds{(e.g., adjusting to a fixed value)} or relative threshold detection \ds{(e.g., adjusting to match a target stimulus)}. This method works well for \ds{target tasks} measured along continuous levels and can provide highly sensitive results \ds{using a relatively small number of stimuli} from a precise sampling of errors across viewers, resulting in concrete, numerical guidelines such as the grid-line alpha values in \cite{bartram2010whisper}. % For example, people may adjust the transparency of a grid until it appears to move in front of a set of marks \cite{bartram2010whisper}. 
	
\vspace{4pt}\noindent\textit{\textbf{Limitations:}}
\label{muscle}
	Experiments using MoA require complex design and implementation. Researchers must choose the levels of target \ds{variable} to test, as well as the starting distances between adjustable \ds{properties of the display} and target stimuli both from-above and from-below. \ds{MoA studies are also sensitive to how people interact with visualizations during the experiment. For example, some studies leverage keyboard inputs (e.g., using the arrow keys to increase or decrease a value) or sliders; however, the design of these inputs may affect how precisely viewers adjust a visualization \cite{matejka2016effect}.} Finally, viewers may develop a motor pattern of adjustment (i.e., ``muscle memory'') that biases their responses over time, sometimes using arbitrary heuristics like, ``10 presses increasing the lightness should be enough.'' To prevent these kinds of predictions and habits, researchers can jitter the adjustment values non-linearly so that one increase or decrease increment is not the same value as the next. Clear task instructions, practice trials, and comprehension checks can help ensure viewer task understanding.

\subsection{Method of Constant Stimuli [MoCS]}
	The Method of Constant Stimuli is among the most common methods in modern psychophysics experiments. Like MoL, MoCS is optimized for detection paradigms (\S \ref{detection}) and is also commonly used for classification, recognition, or identification. MoCS presents viewers with random levels of a target property, presented randomly across trials, and asks them to \ds{draw inferences about that property. Following the previous scatterplot example, researchers can present scatterplots with marks at different lightness levels} and ask viewers to indicate when the dots are present or absent. This method is often used so viewers can assess different properties of the data or display, such as determining which feature (e.g., color or shape) influences average estimation in a scatterplot \cite{gleicher2013perception}.
	
\vspace{4pt}\noindent\textit{\textbf{Advantages:}}
	MoCS enables a diverse range of response types: viewers can be asked to detect an absolute threshold (e.g., “present”/”absent”), much like the method of limits, but they can also be asked to identify relative thresholds based on exemplars (e.g., “greater than x”) or even perform stimuli classification (e.g., “red/blue/green”). Responses can be recorded as binary responses (yes/no), along a continuous (e.g., a magnitude from 0-100), or on a categorical scale (e.g., a color category). MoCS also allows the researcher full control over how the stimuli are sampled (e.g., how wide of a difficulty range is used) to afford creating a full psychometric function or \cn{testing} large response space. These experiments generally provide greater response precision and objectivity due to less viewer habituation: randomizing the order of stimulus levels and interleaving trials with different properties of interest can prevent trial-to-trial response predictions.
	
\vspace{4pt}\noindent\textit{\textbf{Limitations:}}
	Experiments using this method can be complicated to implement. \ds{Because the target property is sampled rather than estimated by the viewer,} it requires a greater number of trials per viewer than other methods. \ds{Researchers must also decide how to sample the space of possible datasets and visualization designs.} This can be modeled through psychometric functions with Ideal Observer Analysis \cite{geisler2003ideal, sims_ideal_2012}. Researchers need to decide and justify how broadly and evenly they sample across variable levels in MoCS experiments. 

\subsection{Adaptive Psychophysical Procedures [APP]}
\label{sec:app}
Because perception is neither perfect nor absolutely precise, viewers may never detect a given magnitude of a stimulus 100\% of the time \cite{kale2018hypothetical}. APPs help researchers find absolute, intensive thresholds in perception \ds{by adapting the stimulus level sampling procedure used in the above methods based on viewer responses}. Researchers using APPs can adjust the visualizations presented to viewers based on their current performance relative to a threshold (e.g., 75\% correct detection) to capture and represent the perceptual processes being measured \cite{cornsweet_staircrase-method_1962}.

The most common APP is staircasing, where experiments increase or decrease the discriminability of presented stimuli depending on the viewer response in the current trial. For example, Rensink \& Baldridge \cite{rensink_perception_2010} use staircasing to find correlation JNDs in scatterplots. They asked viewers to indicate which of two scatterplots has a higher correlation. If viewers respond correctly, the next pair of plots would have closer correlation values; if they respond incorrectly, the next pair would have a larger difference in their correlations. In staircasing, the adjustment continues until some steady-state criteria is met (e.g., 50\% accuracy over $n$ trials). Several algorithmic variants of staircasing and similar techniques are reviewed in detail by Otto \& Weinzierl \cite{otto_comparative_2009}. 
	
\vspace{4pt}\noindent\textit{\textbf{Advantages:}} APPs improve \cn{measurement quality}. They allow researchers to collect precise \cn{performance estimates} using \cn{fewer} trials sampled optimally from the 
%space of 
\cn{possible} levels of the target variable.

\vspace{4pt}\noindent\textit{\textbf{Limitations:}} APPs can be difficult to implement, and researchers must justify their choice in algorithm as well as their choice in steady-state criteria. They also result in varied experiment \cn{duration} that is viewer performance dependent.

%%%%%%%%%%%%%%%%%%%%%%%%%%%%%%%%%%%%%%%%%%%%%%%%%%

\section{Response Types}
\begin{figure}[t]
 \centering
 \includegraphics[width=3in]{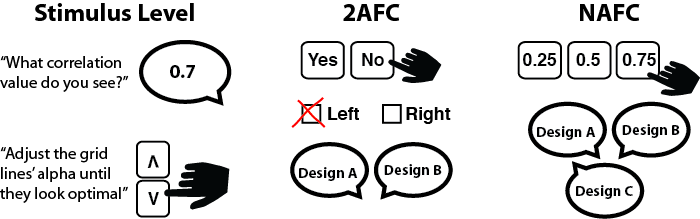}
 \caption{Three ways to elicit responses from viewers during an experiment: stimulus level reporting (left), two-alternative forced-choice (2AFC; center) response, and multiple-alternative forced-choice (NAFC; right).}
 \label{response}
\end{figure}

There are three popular ways to elicit responses from \cn{viewers} during an experiment: stimulus level reporting, two-alternative forced-choice (2AFC) response, multiple-alternative forced-choice (NAFC) (Figure \ref{response}). These response types can be used in a variety of paradigms and can output different dependent measures.

\subsection{Stimulus Level}
Researchers can elicit direct reports, such as perceived values, from viewers. For example, in \cn{Xiong et al.}~\cite{xiong:2019:PerceptualPull}, viewers reported the average vertical positions of lines and bars in a chart by drawing %the new 
\cn{a} line indicating \cn{perceived mean} value on the screen. In \cite{bartram2010whisper}, viewers adjusted the alpha value of gridlines until they considered it to be optimal. Viewers can report stimulus level verbally, by typing in a specific value, or visually by recreating the stimulus level.

\vspace{4pt}\noindent\textit{\textbf{Advantages:}} Stimulus reports enable researchers to measure the specific amount of deviation or bias \cn{of a percept} from ground truth, called error. Because viewers directly report a value instead of selecting from several alternatives, researchers can quantify and model the specific amount viewer reporting deviates from ground truth.

\vspace{4pt}\noindent\textit{\textbf{Limitations:}} Reporting stimulus level can introduce biases like motor inertia (see \S\ref{muscle}) and whole-number bias \cite{hollands2000bias}. For instance, 
\cn{if asked to verbally report scatterplot correlation values}, 
%in a task where viewers verbally report the correlation values of a scatter plot, 
whole-number bias or proportion judgment bias may cause viewers to exclusively report correlation in coarse increments such as 0.25, 0.5, and 0.75.

\subsection{2AFC}
\label{2AFC}
Two-alternative forced-choice tasks give viewers two options after perceiving certain stimuli. Common choices for the two alternatives are comparison and categorization. In 2AFCs designed for comparison, viewers typically 
%choose one of two alternatives to see which measures 
\cn{identify which of two alternatives measures} better on a certain metric. For instance, A/B tests commonly use designs where viewers see two designs and determine which one they prefer. Another type of 2AFC is the “-er” task, where viewers decide which of the two alternatives is [e.g., dark]-``er" than the other. Cleveland \& McGill used this approach 
%in their Graphical Perception \cite{cleveland_graphical_1984} study, where 
\cn{in their canonical study~\cite{cleveland_graphical_1984}, where} 
viewers had to determine which of two \cn{values} were smaller. 

%2AFCs designed for categorization, viewers would determine whether the stimulus they saw met a certain criterion. S\cite{szafir_modeling_2018} adopted this response type where viewers have to respond whether the colors on a scatterplot for two classes were the same or different. \cite{kale2018hypothetical} had viewers view a Hypothetical Outcome Plot and determined whether it portrayed 1) a positive increasing trend or b) a flat trend. 

The choices could be presented verbally or visually. For a verbal task, viewers might be given a series of dashboard interfaces to determine whether each shows an increasing trend in sales. The viewer would indicate yes/no 
%via keyboard response 
upon seeing each dashboard. For a visual manipulation, viewers might be presented with two configurations of a stimulus to choose from. For example, 
%in a usability task to measure interface preference, 
the researcher could present a viewer with two designs of a dashboard and ask them to select the one \cn{showing a greater increasing trend.}
%they prefer more. 

The stimuli and choices in a 2AFC task can be presented over space or over time. In a spatial presentation, viewers see 
%the entirety of the stimuli at the same time 
\cn{both alternatives at once} to make their decision. In a temporal presentation, the stimuli are shown in the same location on the screen but over a certain time interval. For example, to test which interface (A or B) \cn{shows a larger trend}, 
%a user would prefer, before prompting the viewer to choose one, 
a spatial 2AFC design would show both interfaces at the same time, one on the left, and one on the right, while a temporal 2AFC design would show one interface for a certain duration on screen, take it away, then show another interface for a certain duration.
%on the screen. 

\vspace{4pt}\noindent\textit{\textbf{Advantages:}}
\cn{2AFC experiments afford } 
%a 
straightforward data analysis.
%post-experiment. 
The binary response input 
%of yes/no 
allows researchers to classify correct hits, correct rejections, misses, and false alarms. Signal Detection Theory can be used to infer sensitivity and bias \cite{emmerich_signal_1967}, which can in turn help us describe which trends, patterns or visual characteristics are apparent or preferred for a viewer. Another critical advantage of 2AFC tasks is that the researcher can control the rate of criterion they present. 
%For instance, researchers can completely control for criterion effects by “presenting a symmetric, unbiased choice” [25]. This enables the researcher to manipulate any combinations of trade-offs in their investigation. For example, to investigate the trade-off between scattered dot color and size on easiness to perceive in a scatterplot, the researcher could vary the dot color and size incrementally across all the trials in a 2AFC task to identify at which dot size people can no longer differentiate color differences. 
With only two choices, 2AFC tasks also motivate viewers to scrutinize the presented stimuli to capture subtle differences. %This can enhances performance and provides researchers with more information on what drives the difference between the two stimuli. \newline

\vspace{4pt}\noindent\textit{\textbf{Limitations:}} 2AFC tasks may be subject to response bias. When two alternatives are presented to the viewer, they could interact with each other via anchoring effects. Viewers might become more sensitive to the first alternative they see, causing their judgment criteria to change by the time they view the second alternative. %For example, when viewing two scatterplots presented in serial, the separability of the first scatterplot would make the second scatterplot appear more or less separate in comparison. 
Another limitation of the 2AFC task is that it requires multiple viewers or replications of trials to counterbalance stimulus presentation order (to control for order effects). 
%And i
In a preference task, if the two alternatives are equally preferred, the researchers need to aggregate the results of multiple trials and then compare the number of preferences for each alternative to see if they are different. In other words, while it is easy to assess percentage correct in a 2AFC task, obtaining the exact error is difficult without \cn{formally modeling the data.}
%applying modeling to the data. 
%Finally, random guessing is difficult to classify, because there are only two choices. The chance of getting the correct answer is always 50\%. This limitation can be addressed in an NAFC task.

\subsection{NAFC}

NAFC (or N-alternative forced-choice) 
%is a scaled-up version of 
\cn{scales up} a 2AFC task. Instead of presenting the viewer with two alternatives, the researcher shows multiple (N) alternatives. For example, in a correlation classification task, given a ground-truth correlation of 0.5, in a 2AFC task, the researcher might ask the viewer whether the correlation is 0.5 or not, while in a NAFC task, the researcher could ask the viewer to choose the correct correlation from a set of \cn{N values:} 0.25, 0.5, and 0.75 (N=3).
%(here, N = 3). 

\vspace{4pt}\noindent\textit{\textbf{Advantages:}} NAFC tasks %share a lot of the advantages of a 2AFC task, such as easiness to analyze the data. It 
increase how efficiently experiments can detect random guessing. For example, if four alternatives are presented, random chance drops to 25\%. An NAFC task also measures the degree of bias more precisely. For example, %imagine a task where viewers had to determine which bar chart, A or B, has a higher average. In a 2AFC task, they would only report A or B, informing the researcher only which one they perceived to have a higher average without shedding light on how much higher they perceived the average. This can be compensated with a NAFC task. In a NAFC task, 
the researcher could provide the viewer with five options depicting the difference between A and B: A much greater than B, A slightly greater than B, A equals B, A slightly smaller than B, and A much smaller than B.

\vspace{4pt}\noindent\textit{\textbf{Limitations:}} Although NAFC provides more fine-grained information to measure bias, it is still limited by \cn{the size of N.}
%the number of choices the researcher provides for viewers. 
Limitations on human visual attention suggests that viewers should be provided no more than six alternatives \cite{iyengar2000choice}. With six alternatives, it becomes difficult to quantify the specific amount that viewer perception deviates from the ground truth. \cn{As with }
%similar to 
2AFC tasks, while assessing the percentage correct is straightforward, modeling the exact amount of error in response is complicated. Further, the options provided to the viewers could interact with each other or the presented stimulus in memory to cause memory decay, biasing the accuracy of the final response.

%%%%%%%%%%%%%%%%%%%%%%%%%%%%%%%%%%%%%%%%%%%%%%%%%%
%%%%%%%%%%%%%%%%%%%%%%%%%%%%%%%%%%%%%%%%%%%%%%%%%%

\begin{figure}[t]
 \centering
 \includegraphics[width=3.1in]{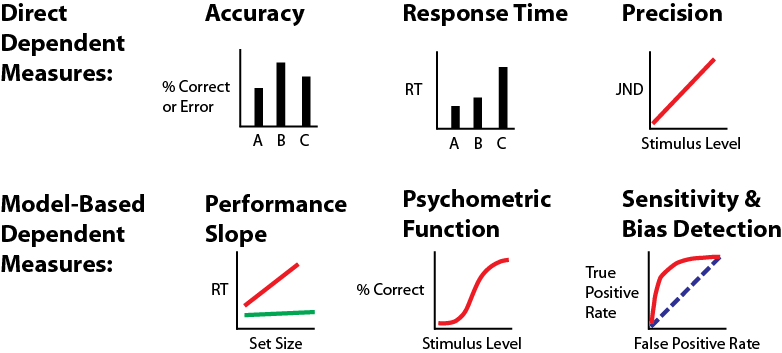}
 \caption{Researchers can analyze dependent measures based on direct measures (top) and model-based measures (bottom). They can analyze viewers' \cn{percentage} correct or degree of error (top left) to understand how close viewers' judgments are to the true value, how quickly viewers can complete a task (top center), variability of viewers' judgments (top right), how robust is a visualization for a given task (bottom left), how decision behavior changes as a function of changes in a stimulus (bottom center), and functions of sensitivity and bias in a ROC curve (bottom right).}
 \label{dependentMeasures}
\end{figure}

\section{Dependent Measures}
\label{sec:dvs}
Dependent measures provide metrics for assessing parameters of a visualization, as shown in Figure \ref{dependentMeasures}. They help researchers concretely quantify how people process visualized data. Experiments should use dependent measures that allow designers to make informed and generalizable decisions about visualizations across a breadth of relevant designs. Once computed, researchers can use a plethora of statistical methods to interpret the resulting outcomes (c.f., \cn{Kay et al.}~\cite{kay2016researcher}). 

\subsection{Direct Dependent Measures}
\label{dep_direct}
We traditionally think of dependent measures as a single number directly measuring how well people process visual information, such as how quickly or how accurately they found a statistic in their data. While visualization largely relies on time and accuracy, efforts such as BELIV have encouraged an expanded library of techniques and measures for assessing visualizations. In vision science, time and accuracy are likewise dominant (though often are only part of the total measure, \S \ref{dep_model}). We refer to measures whose distributions capture performance as direct dependent measures. %These measures offer insight into performance that directly maps onto tasks with minimal translation and assumptions; however, the measures also provide limited generalizability to other data or visualizations. 
Common direct measures include:  

\vspace{3pt}\noindent\textbf{Accuracy: \textit{How close to the true value are people's judgements?}}
%Visualizations traditionally strive to present information in ways that lead to correct conclusions about data. Correctness parallels accuracy, the precision with which a viewer estimates a value. 
Visualization experiments conventionally measure accuracy in two ways: percentage correct (how often do I get the answer right?) and error (how close is my estimate to the true value?). Percentage correct provides a more coarse estimate of visualization effectiveness as a binary correct/incorrect; however, it enables faster responses and greater control over parameters like difficulty. Error offers more precise methods for gauging people’s abilities to infer statistics from data; however, it offers little control over parameters such as task difficulty due to potential confounds and typically requires more time per trial.
 
Accuracy provides an intuitive metric for assessing visualizations, but the simplicity of mean accuracy may hide more sophisticated relationships between visualization design and perception. For instance, while accuracy can tell us whether a value is over- or under-estimated, it cannot tell us how precisely that value is perceived compared to others. Most model-based dependent measures, such as psychometric functions or 
%signal detection 
\cn{sensitivity and bias}, use accuracy to form more nuanced insights into performance.

\vspace{3pt}\noindent\textbf{Response Time: \textit{How quickly can people complete a task?}}
Visualizations typically aim to communicate information both quickly and accurately. Response time (RT) characterizes the time it takes to complete a task with one stimulus. Studies typically use response time either on its own (for simpler tasks) or in conjunction with accuracy (for more challenging tasks \cite{palmer_attentional_1990}) to understand how readily people can infer information from a visualization. While ill-suited for paradigms requiring rapid presentation, RT has often been used in visualization and vision science studies to measure how long it takes people to process visual information with 
%the general consensus that 
\cn{lower response times typically implying} more efficient visual processing.
 
RT provides an intuitive measure well-aligned with traditional visualization goals. However, it requires careful control and, at the time scales of many visualization tasks, may be subject to significant individual differences. As with accuracy, raw RT can inform visualization design; however, it better serves to ground models that allow designers to use experimental outcomes to tailor visualizations to data and tasks. 
 
%\vspace{3pt}\noindent\textbf{Lag: \textit{When do people notice a target?}} 
%While we conventionally think of measuring RT in raw time, 
%Some paradigms allow us to measure RT using other units. Specifically, with paradigms like RSVP (\S \ref{rsvp}), we can measure lag, or the number of elements between a target and a response. %In typical studies using lag, people are asked to identify a target placed at a set position within a series of distractors. %The dependent measure would correspond to the percentage of times the target is correctly identified as a function of the number of distractors preceding the target. For example, in Borkin et al, ...
%Lag allows us to quantify how much information people can attend to at a given time. %It is typically used in studies analyzing memory \cite{} or attending to change \cite{}. 
%In visualizations, this variable helps us to understand how much and how often people can distinguish critical characteristics of a visualization, such as the presence of a particular trend or outlier, especially in dynamic visualizations. While lag gives a measure of RT robust to the input mechanism (e.g., the time to respond is not part of the measure), it also primarily applies to paradigms where stimuli are presented in rapid succession and to research questions involving memory or dynamic visualizations. 
 
\vspace{3pt}\noindent\textbf{Precision: \textit{How variable are people's judgements?}}
Precision is typically quantified as a threshold of performance. Threshold measures correspond to the bounds within which we can reliably observe a particular property of a visualization, such as the level of brightness at which dots in a scatterplot are visible. Studies typically measure thresholds through either adjustment tasks (\S\ref{moa}) where thresholds are derived from the range of values provided by viewers \cite{rensink_perception_2010} or classification tasks (\S\ref{classification}) where thresholds are determined by adjusting parameters of the data or visualization until a desired effect is observed \cite{farell_psychophysical_1999}. One of the most common threshold measures is a just noticeable difference (JND). JNDs are the threshold at which people can detect a difference with a given reliability (typically 50\% or 75\%). 

Thresholds provide probabilistic bounds on the resolution of what people can see in a visualization, allowing designers to reason about how much information is communicated (e.g., the ability to discern the height of a bar or the level of correlation). However, threshold tasks require a significant number of trials 
%to reliably generate 
\cn{due} to the amount of expected noise in \cn{viewers'} responses and the parameter adjustment required to precisely estimate thresholds. 

\subsection{Model-Based Dependent Measures}
\label{dep_model}
We can use factors like time and accuracy to model viewer behavior. %as a function of a visualization’s design or of the underlying data. 
We build these models by aggregating accuracy or RT according to systematically-varied attributes of the visualization or data. We can then use statistical comparisons to draw conclusions from the models. These models allow us to make more precise claims about how the visual system processes information but typically require more data and a more complex experimental design. Model-based measures include: 
 
\vspace{3pt}\noindent\textbf{Performance Slope: \textit{How robust is the visualization for a given task?}}
One method of using accuracy or RT to provide more nuanced insight into a visualization is by modeling how these measures change as a function of the difficulty of a task. A common example of this in vision science is in visual search (\S\ref{search}): how quickly and accurately can I locate a target as the number of targets increases? Experiments measure search slope by systematically varying the number of data points (or another aspect of task difficulty) and tracking accuracy and/or RT at each level. A linear regression would fit a line to the resulting pattern, and the resulting slopes correspond to how sensitive performance is to changes in difficulty. Lower slopes corresponding to more robust designs and a slope of 0 indicates that people can do the task robust to the chosen difficulty level (e.g., finding a point that ``pops-out’’ \cite{wolfe_guided_1994}).

Slopes give us quantitative insight into the relationship between the data and the design by measuring how performance changes with different data characteristics. However, they also assume that performance changes linearly with difficulty and measures changes in RT and accuracy, emphasizing robustness over overall performance. For example, a bar chart may offer lower RT slopes than a dot plot for finding the largest value but only because the bar chart aggregates away the largest value, making it impossible to find.

\vspace{3pt}\noindent\textbf{Psychometric Function: \textit{How does performance change as a function of design?}}
%Certain elements of a visualization’s design, such as the magnitude of a color difference or the proportion of data aggregated, change the conclusions people draw with a visualization. As with task, we can measure change in this behavior as a function of the design of the visualization using psychometric functions. 
Psychometric functions model change in decision behavior (e.g., the number of correct or incorrect decisions) as a function of continuous changes in a stimulus (e.g., the luminance range of a colormap) using a logistic function \cite{farell_psychophysical_1999}. The magnitude of the function at a given point estimates performance for that design setting whereas the spread of the function correlates with noise imparted during the task and the inflection point corresponds to key decision making thresholds. For example, Kale et al. \cite{kale2018hypothetical} use psychometric functions to compare JNDs and noise in trend estimation in uncertainty visualization. 

Psychometric functions offer a way to model decision making behavior as a function of data and design. They capture values at which people predictably make a decision (e.g., when we can reliably estimate which of two marks is larger \cite{smart2019color}?) and the noisy space between where behavior is less well-defined (e.g., how frequently a mark will be estimated as larger?). While psychometric functions offer a powerful tool for modeling perceived values and decision making behaviors, they also require careful experimental control to correctly map the relationship between relevant aspects of a visualization and require tasks that can reliably be modeled as a binary decision (e.g., 2AFC).

\vspace{3pt}\noindent\textbf{Sensitivity \& Bias Detection: \textit{How well does what we see match our data and bias our decisions?}}
\label{signal}
Signal detection \cn{theory} provides a method for modeling performance as a function of sensitivity (how does performance change as the data or design changes?) and bias (do viewers have a tendency towards a certain response?). Signal detection begins by identifying true and false positive and negative responses at different levels of a target independent variable. Once these responses are computed, the resulting patterns are modeled to construct curves showing how sensitivity changes over the corresponding variables, typically with one curve per level of categorical independent variable. \cn{Bias corresponds} to the curve intercepts, and sensitivity \cn{corresponds} to the parameters of the curve. Signal detection is explained in more detail in other sources \cite{emmerich_signal_1967,geisler_sequential_1989,geisler2003ideal,mcnicol_primer_2005,stanislaw_calculation_1999}.
%Emmerich \cite{emmerich_signal_1967}; Geisler \cite{geisler_sequential_1989, geisler_ideal_2002}; McNicol \cite{mcnicol_primer_2005}; and Stanislaw \& Todorov \cite{stanislaw_calculation_1999}. 

As with psychometric functions, 
%signal detection 
\cn{sensitivity and bias detection} allow us to measure how well a visualization performs under different conditions and statistically disentangles meaningful aspects of performance from noise. It also allows researchers to measure potential bias in visualization interpretation and can apply to tasks beyond conventional decision making and detection \cn{as well as those} that may not have a linear or logistic pattern. \cn{Sensitivity and bias detection} enable researchers to use more traditional algorithmic analyses, such as ROC curves \cite{stanislaw_calculation_1999}, to analyze their results. However, like other model-based measures, 
%signal detection requires 
\cn{using these measures requires} experiments that are carefully structured to systematically manipulate relevant variables, and comparing sensitivity parameters is a level of abstraction removed from more direct metrics like accuracy or slope. 

\section{Discussion \& Future Work}
In this paper, we discuss the value of using perceptual methods in rigorous visualization experiments. Our design space can be applied to facilitate novel research by systematically examining viewer behavior and to produce replicable and generalizable design guidelines. This paper is non-exhaustive and prioritizes breadth and structure of methods over depth. This is not a complete handbook on how to study perceptual mechanisms; however, we do anticipate our design space being highly useful for experimenters, reviewers, and readers in critical planning of new studies and evaluation of past work.

We cover task design topics extensively but exclude fundamentals of behavioral research such as experimental control, the basics of hypothesis testing, implementation (e.g., experiment software), and materials (e.g., hardware). Additionally, essential modeling techniques such as Signal Detection Theory, Ideal Observer Analysis, and the statistical methods used to analyze experimental data (e.g., Bayesian inference) are beyond the scope of this work. We strongly encourage researchers to consider these topics as part of any experimental design.
%familiarize themselves with these topics.

Visualization research can sample this design space to construct experiments that measure key components of visualization design and help bridge findings from vision science. \cx{We have included a supplementary guide showcasing how an experiment could be designed following this design space
%, see 
\cn{at} \href{https://visxvision.com/using-the-design-space/}{https://visxvision.com/using-the-design-space/}. Note that while experiments can use these methods in common configurations (Fig. \ref{structureOverview}), we hope that the structure provided by these four phases will also yield novel and innovative studies}. Shared methodologies and associated vocabularies can help researchers understand what makes visualization comprehension unique from other visual experiences. For example, these methods may explore critical thinking affordances in visualizations or biases from visual illusions. Understanding visualization as a function of design and of the visual mechanisms used to process those designs may lead to broadly generalizable guidelines and more effective visualization practices. By providing a library of common techniques and relevant terminology, we hope to bridge lexical divides between visualization and vision science and better facilitate these innovations.
%inquiry. 

%While our focus is on how these methods can support visualization, vision science could benefit from the breadth of cognitive and perceptual questions involved in interpreting visualizations. Future work should expand upon our design space to help facilitate research towards more obscure aspects of perception and cognition such as visual illusions, critical thinking and learning.  By exploring common experimental techniques and introducing key terms critical for these techniques, we hope this work will help bridge lexical divides between visualization and vision science to facilitate collaboration and interdisciplinary investigation.

%However, visualization research offers opportunities to explore innovative 

%Connecting these methods with high-level task goals and perceptual mechanisms might 
%\todo{wanted to add more of a synthesis component. check this} 
%also be useful. Finally, we encourage authors to adopt the terms used in this paper to describe their experiments and help standardize methodological language in visualization research. 

\section{Conclusion}
We provide a design space of vision science methods for visualization research, along with a \cn{shared} lexicon for facilitating deeper conversation between researchers in both fields. Our goal was to synthesize and organize the most popular experimental tools to provide a foundation for evaluating and conducting future research. We hope that the visualization community sees the value in diversifying experimentation, and embracing new approaches to advance our knowledge about both human behavior and guidelines for design.

\acknowledgments{
We thank Dr. Ronald Rensink for his feedback and suggestions and Dr. Tamara Munzner, Dr. Juergen Bernard, Zipeng Liu, Michael Oppermann, Francis Nguyen, and our reviewers for their thoughtful comments. This work was supported by
NSF \#1764092, \#1764089, \#1657599, and the UBC 4 Year Ph.D. Fellowship.}

\balance
\bibliographystyle{abbrv-doi}

\bibliography{MyLibrary.bib}
\end{document}